\documentclass[iop,apj,twocolumn,a4paper]{aastex631}
\usepackage{amsmath,amstext}
\usepackage{hyperref}
\usepackage{amssymb}
\usepackage{wasysym}
\usepackage{float}
\usepackage{ulem}
\usepackage{breakurl}
\usepackage{color}
\usepackage[flushleft]{threeparttable}

\def\msun{\,{\rm M_\odot}}

\def\spose#1{\hbox to 0pt{#1\hss}}
\def\lta{\mathrel{\spose{\lower 3pt\hbox{$\mathchar"218$}}
     \raise 2.0pt\hbox{$\mathchar"13C$}}}
\def\gta{\mathrel{\spose{\lower 3pt\hbox{$\mathchar"218$}}
     \raise 2.0pt\hbox{$\mathchar"13E$}}}

\lefthead{Madau}
\righthead{X-ray Weak Super-Eddington AGNs}
\journalinfo{}


\begin{document}

\title{X-Ray Weak AGNs from  Super-Eddington Accretion onto Infant Black Holes}

\author[0000-0002-6336-3293]{Piero Madau}
\affiliation{Department of Astronomy \& Astrophysics, University of California, 1156 High Street, Santa Cruz, CA 95064}
\affiliation{Dipartimento di Fisica ``G. Occhialini", Università degli Studi di Milano-Bicocca, Piazza della Scienza 3, I-20126 Milano, Italy}

\author{Francesco Haardt}
\affiliation{Dipartimento di Scienza e Alta Tecnologia, Universit\`a degli Studi dell'Insubria, via Valleggio 11, I-22100 Como, Italy}
\affiliation{INAF, Osservatorio Astronomico di Brera, Via E. Bianchi 46, I-23807 Merate, Italy}
\affiliation{INFN, Sezione Milano-Bicocca, P.za della Scienza 3, I-20126 Milano, Italy}

\begin{abstract}
A simple model for the X-ray weakness of JWST-selected broad-line AGNs is proposed under the assumption that the majority of these sources are fed at super-Eddington accretion rates. In these conditions, the hot inner corona above the geometrically thin disk that is responsible for the emission of X-rays in ``normal" AGNs will be embedded instead in a funnel-like reflection geometry. The coronal plasma will Compton upscatter optical/UV photons from the underlying thick disk as well as the surrounding funnel walls, and the high soft-photon energy density will cool down the plasma to temperatures in the range 30--40 keV. The resulting X-ray spectra are predicted to be extremely soft, with power-law photon indices $\Gamma\simeq\,$ 2.8--4.0, making high-$z$ super-Eddington AGNs largely undetectable by {\it Chandra}.
\end{abstract}
\keywords{Accretion (14); Active galactic nuclei (16); James Webb Space Telescope (2291); Supermassive black holes (1663)}

\section{Introduction}

Deep surveys with the James Webb Space Telescope (JWST) have revealed an emergent, large population of moderate-luminosity, broad-line active galactic nuclei (AGNs) at $z=4-10$  powered by accretion onto $M=10^{6}-10^8\,\msun$ early massive black holes (MBHs) \citep[see, e.g.,][]{Kocevski2023,Harikane2023AGN, MaiolinoAGN}. These MBHs appear to be overmassive \citep[but see][]{Ananna2024} compared to their galaxy hosts and to grow while being extremely weak in X-rays \citep[e.g.][]{Maiolino2024b,Yue2024}. It has already been suggested that accretion at super-Eddington rates may answer many of the theoretical challenges posed by these and other observations. Some of the attractive features of this scenario include the very rapid grow of light seeds occurring during short-lived supercritical episodes \citep[][]{Madau2014,Volonteri2015,Pezzulli2016}, as well as the artificial bias in black hole mass estimates induced by the anisotropic radiation field emitted by thick accretion tori \citep[e.g.][]{King2024,Lupi2024b}.

Supercritical accretion flows are known to form geometrically and optically thick disks dominated by radiation pressure. The large thickness of the disk naturally collimates radiation and produces a highly super-Eddington photon flux along the rotation axis
\citep[for a review, see][]{Abramowicz2013}. In this regime, the inner, hot and luminous funnel region remains hidden from view, visible only at small viewing angles from the rotation axis of the system \citep[e.g.,][]{Sikora1981,Madau1988,Sadowski2014,Jiang2014, Ogawa2017}. \citet{Pacucci2024} have recently argued that the diminishing X-ray contribution predicted at high inclination angles from the pole may offer an explanation for the X-ray weakness of the ``little red dots", an enigmatic subcomponent of JWST-selected broad-line AGNs \citep[e.g.][]{Greene2024,Kokorev2024, Matthee2024}.

In this Letter we propose a simple model for the X-ray weakness of JWST-selected AGNs. At super-Eddington rates, the hot corona above the  inner accretion disks that is thought to be responsible for the emission of X-rays in ``normal" AGNs \citep[e.g.,][]{Haardt1991} is embedded in a funnel-like reflection geometry. The coronal plasma upscatters soft photons from the underlying disk as well as the surrounding funnel walls, and the high UV energy density cools down the plasma to temperatures below 40 keV. We shall find that the resulting X-ray spectra are extremely soft, with hard X-ray bolometric corrections that can be two orders of magnitude larger than those of standard AGNs.

\section{Supercritical Accretion Flows}
\label{sec:PW}

The shape and luminosity of low-viscosity, rotating, radiation-pressure supported accretion flows around black holes have been computed by several authors under a number of simplifying assumptions \citep[e.g.,][]{Paczynsky1980,Wiita1982,Wielgus2016}. These axisymmetric, $\dot M\gg \dot M_{\rm Edd}$, semi-analytical ``thick disk" models differ from the standard ``slim disk” solution \citep{Abramowicz1988, Sadowski2009,Wang2014,Lasota2016}, in that they are geometrically thick and radiatively efficient. We choose and briefly summarize below the thick disk formulation because numerical simulations of super-Eddington accretion flows have shown that vertical advection of radiation caused by magnetic buoyancy transports energy faster than radiative diffusion \citep[][]{Jiang2014,Jiang2019}. This effect allows photons to escape from the surface of the thick disk before being trapped and radially advected into the hole and undermines the underluminous -- at a given $\dot M$ -- slim disk solution.

Let us adopt a cylindrical coordinate system ($r, \varphi, z)$ centered on a Schwarzschild black hole of mass $M$, and use a pseudo-Newtonian potential to mimic general relativistic effects, $\Phi=-GM/(R-r_S)$, 
where $R=(r^2+z^2)^{1/2}$ is the spherical radius and $r_S=2GM/c^2$ is the  Schwarzschild radius. A physically realistic specific angular momentum distribution $\ell(r)$ of the form
\citep{Paczynsky1980,Wiita1982}
\begin{equation}
\ell(r)=\ell_K(r_{\rm in})+{\cal C}(r-r_{\rm in})
\label{eq:angmom}
\end{equation}
completely determines the shape of a thick disk (in the case of a polytropic gas $\ell$ depends only on the position coordinate $r$). The disk half-thickness $z=h(r)$ can be shown to be given by
\begin{equation}
h(r)=\left\{\left[{GM(r_{\rm in}-r_S)\over GM-(r_{\rm in}-r_S) J(r)}+r_S\right]^2-r^2\right\}^{1/2},
\end{equation}
where $J(r)\equiv \int_{r_{\rm in}}^r\ell^2(r')dr'/r'^{3}$. The intersections of $\ell(r)$ with the Keplerian angular momentum distribution $\ell_K(r)$ provides the inner edge of the torus ($r_{\rm in}$), a pressure maximum ($r_c$), and an outer transition radius ($r_{\rm out}$) where the thick solution matches on to a thin accretion disk. The inner radius $r_{\rm in}$  lies between the marginally bound and marginally stable orbits; {its location determines the efficiency $\varepsilon$ of the conversion of accreted matter into radiation.} 

\begin{figure}[!htb]
\centering
\includegraphics[width=\hsize]{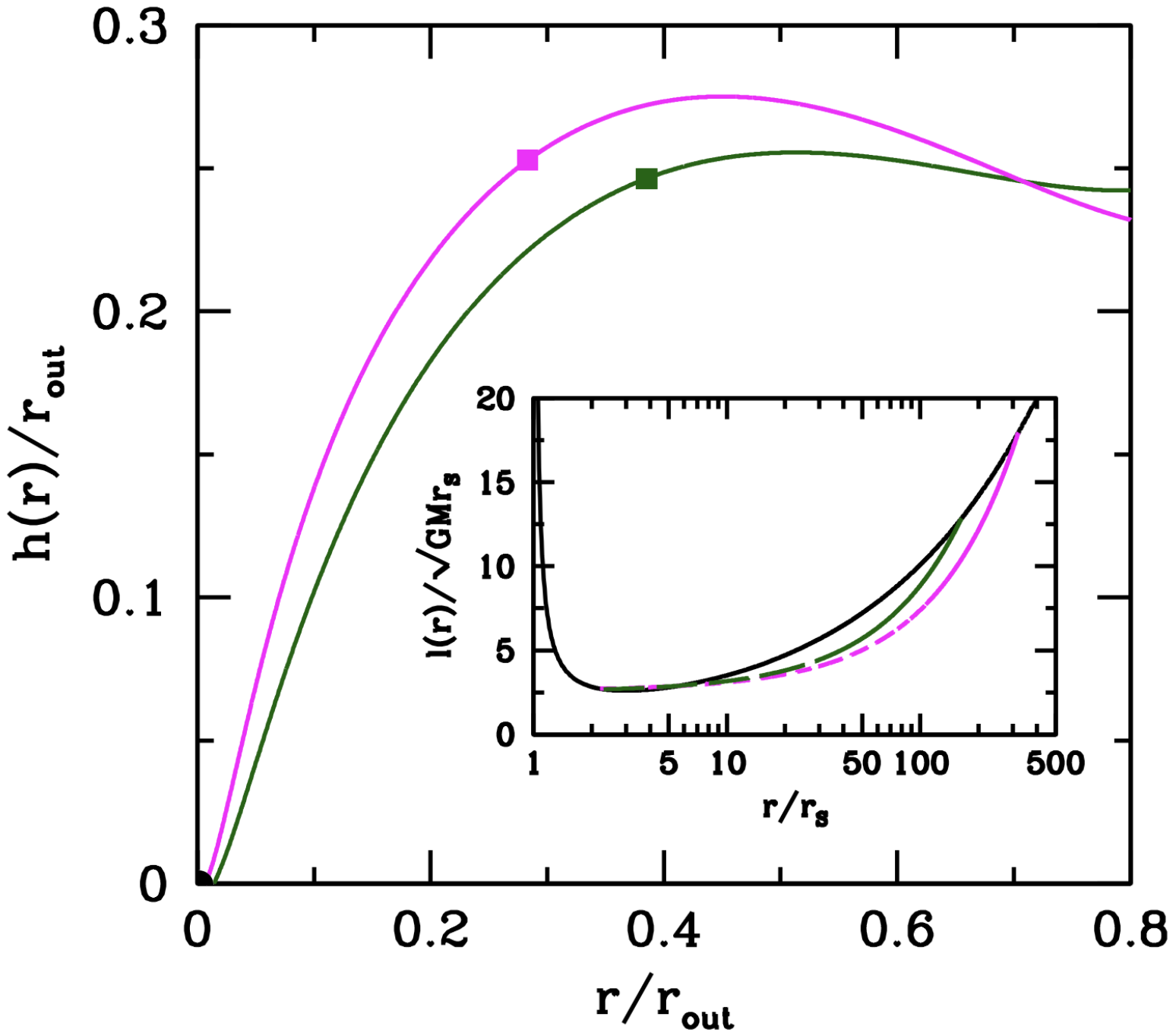}
\caption{Meridional cross-sections (over one quadrant) for the supercritical thick disks described by Models A (magenta line) and B (green line). As $r_{\rm in}$ decreases, $r_{\rm out}$ increases as does the ratio $L/L_{\rm Edd}$, while  the efficiency of mass to energy conversion drops. Smaller values of $r_{\rm in}$ imply steeper and deeper funnels, where pressure gradients are balanced by centrifugal forces rather than by gravity and luminosities exceed the Eddington limit. The square points mark the location on the surface inside which 90\% of the disk luminosity $L_{\rm rad}$ is actually emitted. The inset shows the Keplerian specific angular momentum distribution
for the adopted pseudo-Newtonian potential (black line), and the angular momentum distributions corresponding to our supercritical disks (Models A and B). 
}
\label{fig:shape}
\end{figure}

Radiation is emitted from the photosphere at the local Eddington rate
\begin{equation}
{\vec F}_{\rm rad}=-{c\over \kappa_{\rm es}}{\vec g}_{\rm eff}=
-{c\over \kappa_{\rm es}}\left(-{\vec \nabla}\Phi+{\ell^2\over r^3} {\hat e_r}\right), 
\label{eq:Frad}
\end{equation}
where ${\vec g}_{\rm eff}$ is the effective gravity vector  perpendicular to the surface of the disk, and $\kappa_{\rm es}$ is the electron  scattering opacity. The total luminosity radiated by the torus, $L_{\rm rad}$, is calculated by integrating the emitted flux over the two disk surfaces,
\begin{equation}
\begin{aligned}
L_{\rm rad} & = 2\int_{0}^{2\pi}\int_{r_{\rm in}}^{r_{\rm out}}F_{\rm rad}\,d\Sigma,
\label{eq:Lrad}
\end{aligned}
\end{equation}
where $F_{\rm rad}\equiv |{\vec F}_{\rm rad}|$ and $d\Sigma=[1+(dh/dr)^2]^{1/2}\,rdrd\varphi$ is the disk area element. In the case of large tori, the total luminosity generated by viscosity is simply given by 
\begin{equation}
L_{\rm gen}=\dot M e_{\rm in}=\dot M c^2\, {r_S(r_{\rm in}-2r_S)\over 4(r_{\rm in}-r_S)^2}\equiv \varepsilon \dot M c^2,
\label{eq:Lgen}
\end{equation}
where $e_{\rm in}$ is the specific binding energy at the inner edge. Global energy conservation requires that the total energy gain be compensated by radiative losses, i.e. $L_{\rm gen}=L_{\rm rad}$. By equating Equations (\ref{eq:Lrad}) and (\ref{eq:Lgen}) we then finally calculate $\dot M$.

\begin{table}
  \begin{threeparttable}
    \caption{Properties of Thick Accretion Disks. \label{table:thickdisks}}
     \begin{tabular}{cc}
        \toprule
        Parameter & Values for Models A and B\\ [2pt]
        \hline
        $r_{\rm in}/r_S$  & 2.2, 2.3 \\
        ${\cal C}$  & 0.04792, 0.063301\\  
        $r_{\rm out}/r_S$ &  320, 163\\
        $\varepsilon$ &  0.035, 0.044\\
        $L_{\rm rad}/L_{\rm Edd}$ &  4.3, 3.1\\
        $\dot M/\dot M_{\rm Edd}$ &  12.5, 7.1\\
        $\Theta$  & 35$^\circ$, 44$^\circ$\\ 
        \hline
     \end{tabular}
  \end{threeparttable}
\end{table}

\begin{figure}[!hbt]
\centering
\includegraphics[width=\hsize]{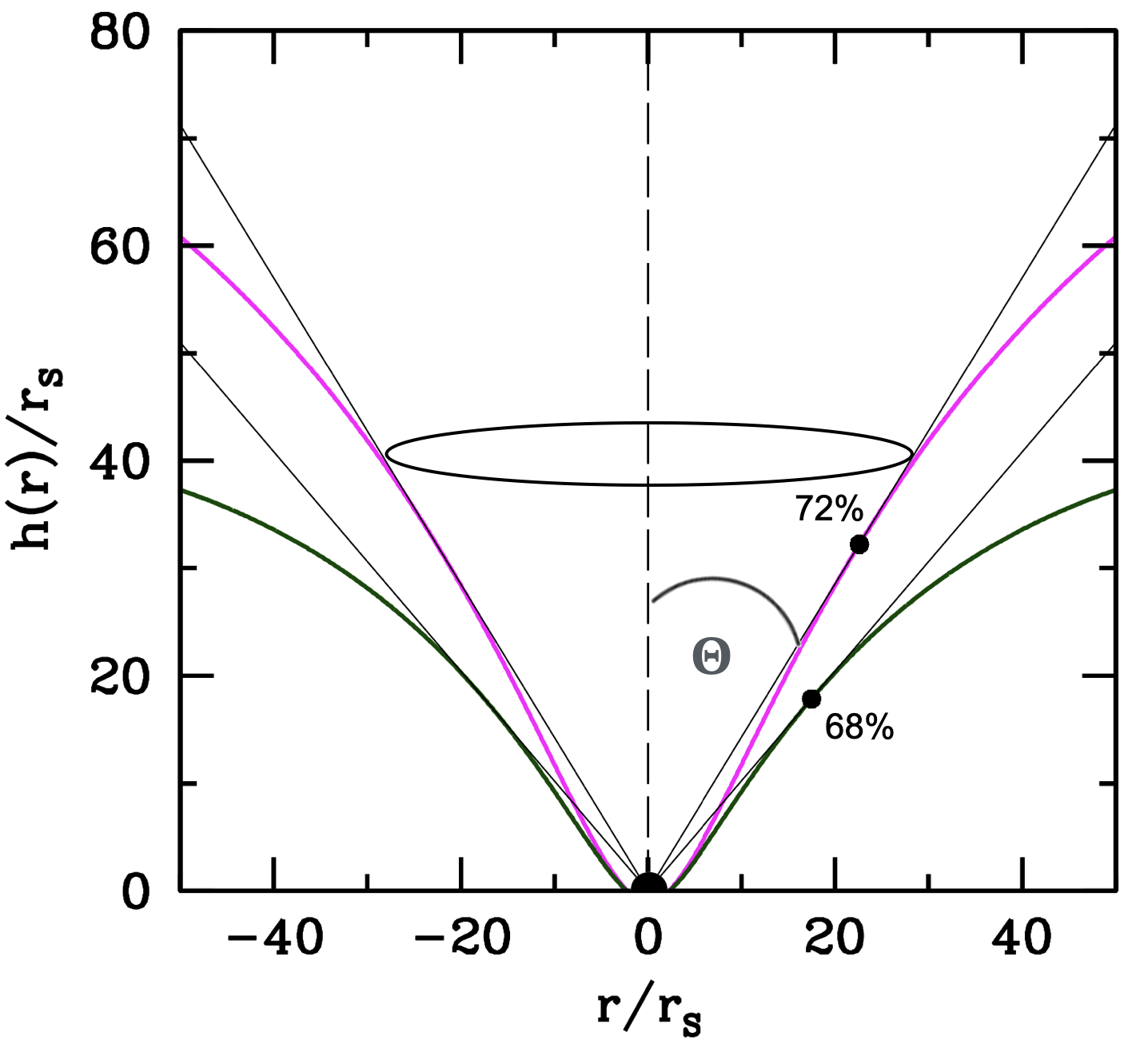}
\caption{The funnel regions of supercritical disks around black holes. The two configurations (Models A and B)  have funnel opening half-angles of $\Theta={\rm arccot}\,(h/r)_{\rm max}=
35^\circ$ and $44^\circ $, respectively. The dot on each funnel indicates the point where $h/r$ reaches a maximum, and the percentage next to it denotes the fraction of total disk luminosity
emitted interior to that point.
}
\label{fig:funnel}
\end{figure}

The properties of two representative super-Eddington disk models are summarized in Table \ref{table:thickdisks}. Values for the disk inner edge $r_{\rm in}$ and the constant ${\cal  C}$ in Equation (\ref{eq:angmom})  were chosen following \citet{Paczynsky1980}.  As $r_{\rm in}$ decreases, the disk grows larger and fatter (narrower funnel) and the accretion flow becomes more super-Eddington. The locally-generated radiation flux peaks at a few Schwarzschild radii and then drops as $\sim r^{-2}$ in the inner funnel. The total isotropic luminosity radiated in Models A and  B is $L_{\rm rad}/L_{\rm Edd}=4.3$ and 3.1, corresponding to $\dot M/\dot M_{\rm Edd}=12.5$ and 7.1, respectively. Here, we have defined the critical accretion rate, $\dot M_{\rm Edd} \equiv 10\,L_{\rm Edd}/c^2$, for powering the Eddington luminosity assuming a 10\% radiative efficiency. The shape of the thick portion of the disks (one quadrant only) is outlined in Figure \ref{fig:shape}, while Figure \ref{fig:funnel} depicts the narrow, low-density funnel that develops in the innermost regions. This can be  characterized by the opening half-angle $\Theta$: 
\begin{equation}
\Theta={\rm arccot}\,(h/r)_{\rm max}.
\end{equation}
Soft photons emitted within the funnel  scatter off its  walls many times before escaping to infinity. 

\section{Coronal X-ray Emission}
\label{sec:XR}

The X-ray emission of AGNs is thought to be generated by thermal Comptonization of UV photons by hot electrons in a corona \citep[e.g.,][]{Haardt1991}. The soft seed photons originate in a cold, geometrically thin accretion disk. The geometry of the hot X-ray corona  is poorly known, but a variety of observational constraints point to a very compact X-ray source located within $\lta 10$ Schwarzschild radii of the black hole \citep[e.g.,][]{Reis2013,Fabian2015}. In general, the hard X-ray spectra of Seyfert galaxies can be well approximated by a power law with photon index $\Gamma= 1.7–2.0$ and a high-energy cutoff. These  depend on the physical properties of the scattering corona, i.e., the optical depth $\tau$ and the electron temperature $kT_e$
\citep[e.g.,][]{Tortosa2018}.

To extend this two-phase disk-corona scenario to supercritical flows, we place an artificial, plane-parallel, thin corona just above the funnel inner surface, and denote with $f_c$ the fraction of the locally generated radiation power that emerges from this hot tenuous phase. Each area element within the funnel will ``see" incoming radiation that must be included in the balance of forces. Equation (\ref{eq:Frad}) must then be modified as:
\begin{equation}
{\vec F}_{\rm rad}+{\vec F}^n_{\rm in}=-{c\over \kappa_{\rm es}}{\vec g}_{\rm eff},
\label{eq:Ffun}
\end{equation}
where now it is the net flux $\vec F_{\rm rad}+{\vec F}^n_{\rm in}$ that balances gravity. Note that  
${\vec F}_{\rm rad}$ and  the normal component of the incoming flux ${\vec F}^n_{\rm in}$ point in opposite directions.

Consider first, for illustration, a geometrically thin disk scenario with no incoming flux, and let us write the total Compton luminosity per unit area in the corona as the sum of an upward-, $F_c^+$, and downward-directed, $F_c^-$, component. Ignoring the anisotropy of inverse Compton scattering, we can express the power radiated  in all directions by the hot phase as $2F_c^+=ApF_s$, 
where $A$ is the mean energy gain per scattering, $F_s$ is the total soft photon flux emerging from the cool disk, and $p\approx 1-\exp(-2\tau)$ is the mean probability of scattering in a plane-parallel corona of total optical depth
$\tau$ measured in the vertical direction. The outgoing photon flux is $F_{\rm rad}=F_s+F_c^+$. Hard X-ray photons emitted by the corona and directed downward toward the cool, optically thick disk layers are largely absorbed except for a small reflected component, \citep{Lightman1988}. Under the assumption of a nonreflective disk, energy balance for the cold and hot phases then gives
\begin{equation}
\begin{aligned}
F_s=& (1-f_c)\,{c\over \kappa_{\rm es}}g_{\rm eff}+{1\over 2}ApF_s,\\
F_c^+& ={f_c\over 2}\,{c\over \kappa_{\rm es}}g_{\rm eff}.
\end{aligned}
\label{eq:balance1}
\end{equation}
In the top equation, the first term represents the primary soft photon flux diffusing from below, while the second term corresponds to the X-ray power radiated by the corona toward the disk surface, where it is absorbed and thermalized. The bottom equation expresses the upward Compton X-ray flux, $F_c^+$, as a fraction $f_c/2$ of the total generated power.
Solving for $A$ and using the previous definitions, we derive 
\begin{equation}
A={2f_c\over (2-f_c)p}.
\label{eq:Amp}
\end{equation}
In the regime of unsaturated Comptonization, the mean energy gain per scattering is given by $A=4\theta+16\theta^2$, where $\theta=kT_e/m_ec^2$ is the dimensionless electron temperature. Together with Equation (\ref{eq:Amp}), this implies a relation between the luminosity of the corona and its scattering optical depth and temperature,  and therefore the photon index $\Gamma$ of the  power-law Comptonized component $F_\nu\propto \nu^{-\Gamma+1}$ \citep{Haardt1993}. The Comptonization radiative transfer models of \citet{Sunyaev1980} give
\begin{equation}
\Gamma={-\ln\tau+2/(3+\theta)\over \ln(12\theta^2+25\theta)}+1.
\label{eq:alpha}
\end{equation}
%
%
%
%
%
%
According to the above expressions, spectral indeces and electron temperatures are in the range $\Gamma=1.70-2.0$ and $kT_e\gta 70\,$ keV for $0.3\lta \tau\lta 0.8$ when one third of the soft  photospheric emission of the disk results from the reprocessing of the hard X-ray photons emitted by the local corona, i.e. when $f_c=0.5$. Such hard X-ray photon indeces are seen above 2 keV in low-redshift radio-quiet AGNs \citep[$\Gamma=1.89\pm 0.11$;][]{Piconcelli2005}. The highest luminosity AGNs at $z\gta 6$ have X-ray spectra that are steeper than these values  \citep[$\Gamma=2.4\pm 0.1$;][]{Zappacosta2023}, and so do near-Eddington  narrow-line Seyfert 1 (NLS1) galaxies \citep[$\Gamma=2.68\pm 0.51$;][]{Grupe2010}. Note that the thin disk-corona scenario is highly anisotropic, as soft  photons enter the corona only from below.

\begin{figure}[!hbt]
\centering
\includegraphics[width=\hsize]{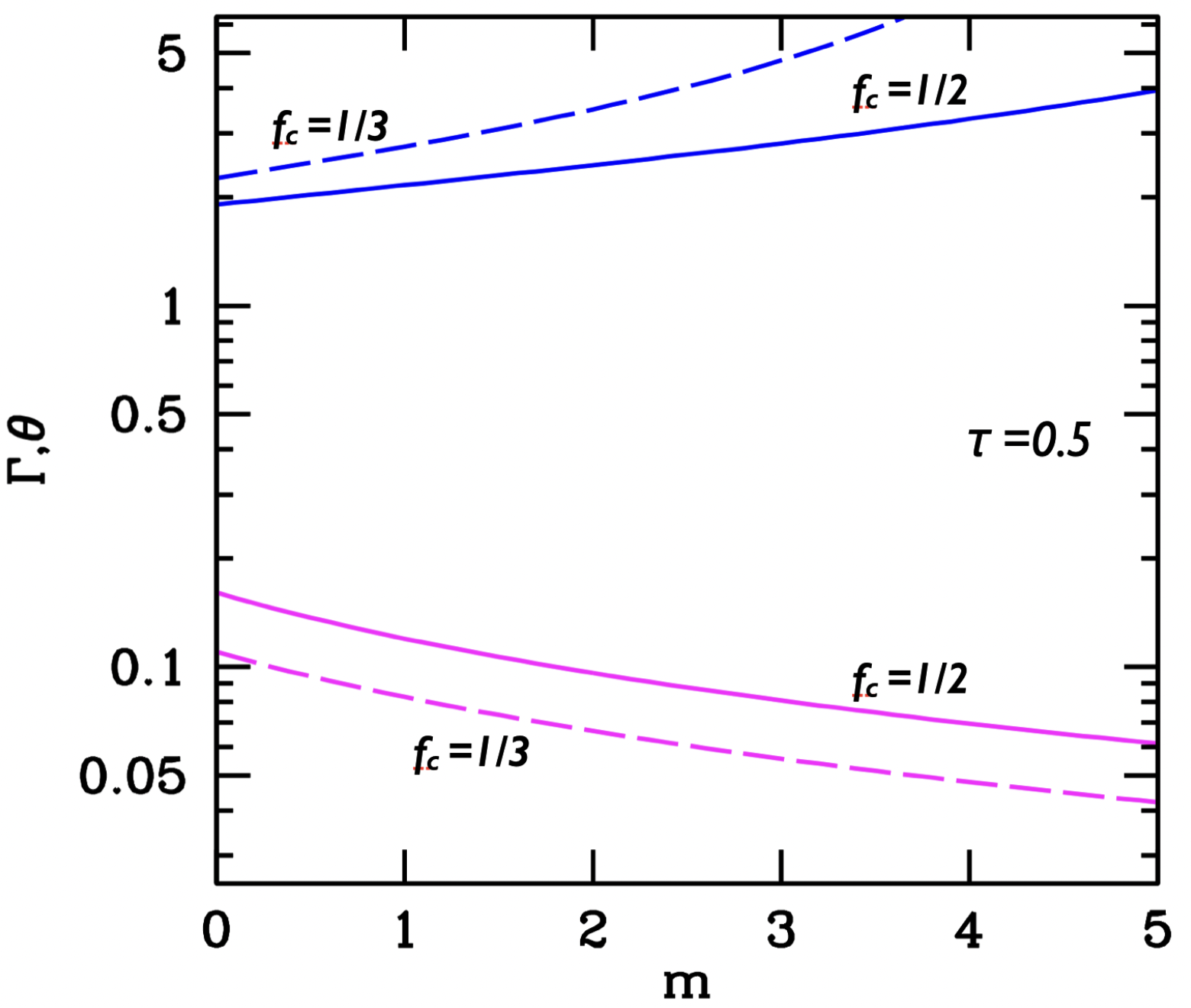}
\caption{Equilibrium electron temperature ($\theta=kT_e/m_ec^2$) of the hot coronal phase (lower curves) and the resulting photon index $\Gamma$ of the X-ray Comptonized component (upper curves). The corona has an assumed electron scattering opacity of $\tau=0.5$, comparable to the value inferred in  Seyfert galaxies. Two values, $f_c=1/2$ and $f_c=1/3$, have been assumed for the fraction of the locally-generated radiation power that is dissipated in the corona. The parameter $m$ on the horizontal axis measures the strength of the incoming external radiation field in units of the net flux. In our two supercritical thick disk models, we estimate $m\simeq 3$ (Model A) and $m\simeq 2$ (Model B) at the bottom of the funnel.
}
\label{fig:Xray}
\end{figure}

\begin{figure}[!hbt]
\centering
\includegraphics[width=\hsize]{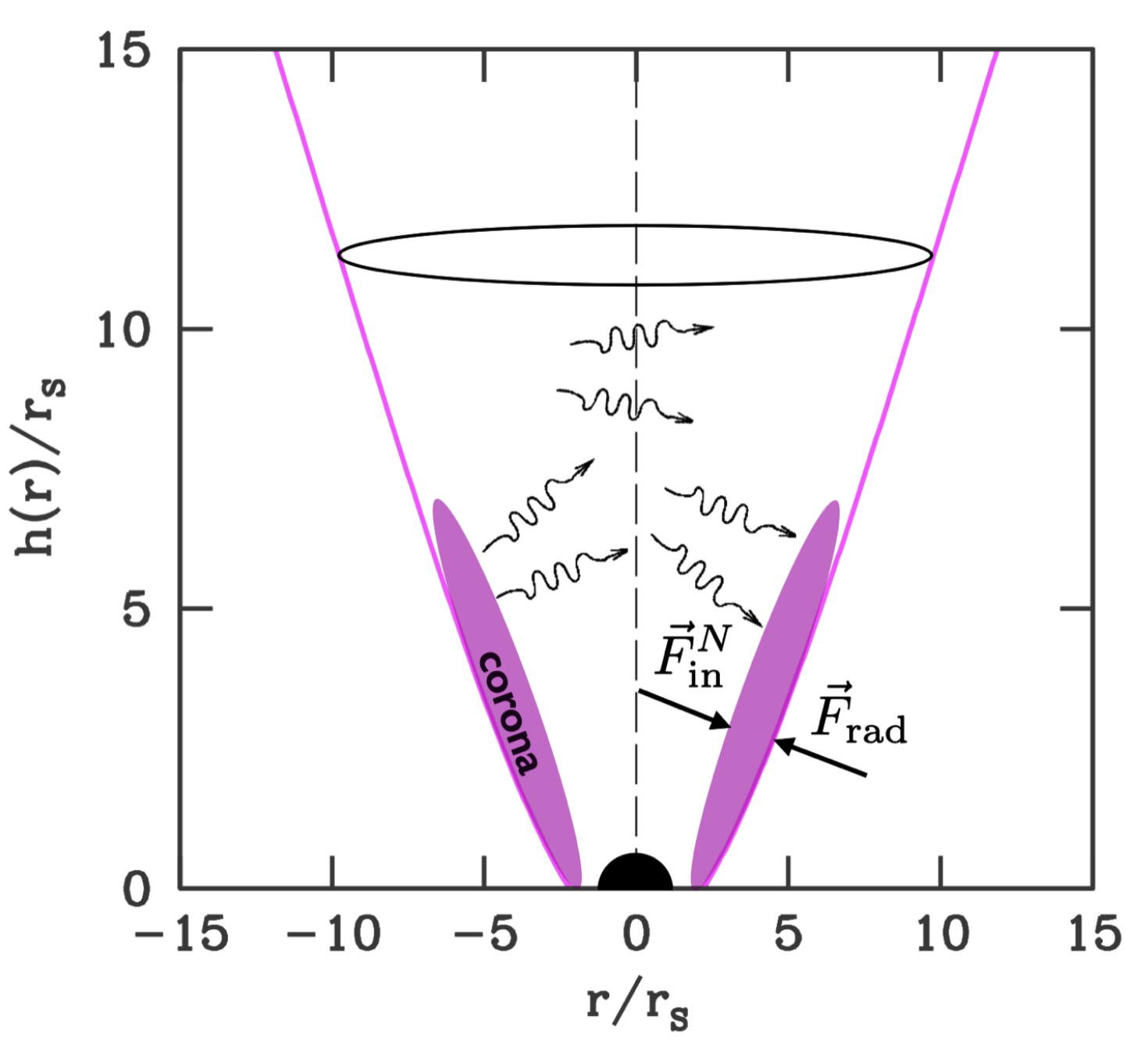}
\caption{Sketch illustrating the super-Eddington corona-funnel model discussed in the main text. Embedded in a funnel-like reflection geometry, the inner hot corona will cool down by Comptonizing a largely isotropic soft radiation field. The resulting X-ray spectra are predicted to be extremely soft, independently of the detailed geometry of the corona.
}
\label{fig:corona}
\end{figure}

Consider instead the case of a thick supercritical disk. 
Near the bottom of the narrow funnel, where much of the luminosity is generated, the soft radiation field has a large isotropic component. The coronal plasma will Compton cool not just on the locally-generated soft photons coming from below, but also on those emitted by the surrounding funnel walls and entering the corona from above. The temperature of the hot electrons will drop, and the emitted X-ray spectrum will be much softer than derived previously. This can be seen by rewriting Equations (\ref{eq:balance1}) as
\begin{equation}
\begin{aligned}
F_s=& (1-f_c)\,{c\over \kappa_{\rm es}}g_{\rm eff}+{1\over 2}ApF_s+
m {c\over \kappa_{\rm es}}g_{\rm eff}(1-p),\\
F_c^+& ={f_c\over 2}\,{c\over \kappa_{\rm es}}g_{\rm eff},
\end{aligned}
\label{eq:balance2}
\end{equation}
where the third term  represents the incoming external photon flux that is added to the soft photon pool. We have  expressed this term as a multiple $m (1-p)$ of the net flux $(c g_{\rm eff}/\kappa_{\rm es})$, where the factor $1-p$ modulates both the soft photon external flux emerging from the disk$+$corona system on ``the other side of the funnel" as well as the incoming X-ray radiation that is transmitted by the local corona, absorbed, and thermalized. Our formalism assumes that external X-ray photons undergo coherent scattering with local coronal electrons, resulting in no change in photon energy. The validity of this approximation will be explored in a future study. It has long been recognized that in narrow accretion funnels (small half-angle $\Theta$) the net flux ${F}_{\rm rad}-{F}^n_{\rm in}$ is a much smaller quantity than either ${F}_{\rm rad}$ or ${F}^n_{\rm in}$ because of the strong ``reflection effect" \citep{Sikora1981,Narayan1983,Madau1988}. Solving for the amplification factor $A$ as before we now have
\begin{equation}
A={2f_c\over [(2-f_c) +2m(1-p)]p}.
\label{eq:Amp2}
\end{equation}
Figure \ref{fig:Xray} demonstrates how, with increasing values of $m$, AGN X-ray spectra become extremely soft as the corona cools down. Our scenario is schematically illustrated in Figure \ref{fig:corona}. 
To estimate the parameter $m$ in the limit $p=0$, we have subdivided the surface of the funnel into a finite number of $r$-rings with quasi-uniform emission properties, and approximated the force balance Equation (\ref{eq:Ffun}) as
\begin{equation}
F_{\rm rad,i}-{1\over \pi} \sum_j B_{ij}\,F_{\rm rad,j}={c\over \kappa_{\rm es}}{g}_{\rm eff,i},
\end{equation}
where 
\begin{equation}
B_{ij}={{\hat n_i}\cdot \vec D\over |{\vec D}|} \left[
{(\hat n_j\cdot \vec D)\over |\vec D|^3} d\Sigma_j\right]
\end{equation}
is the fraction of radiation emitted from the $j$th ring that reaches the $i$th ring along the direction $\vec D$ in the solid angle given by the term in square brackets. Here, ${\hat n}_j$ is the outward unit normal to the surface at the element $d\Sigma_j=r_j [1+(dh/dr)_j^2]^{1/2}\,d\varphi\Delta r$, and $\Delta r$ is the mesh size. Solving this set of linear equations for Models A and B gives $m\simeq 3$ and $m\simeq 2$ at $r_i=10\,r_S$, respectively.  Models with higher supercritical rates (smaller $\Theta$) yield values of $m>3$ near the bottom of the funnel.

Consider now a ``standard" AGN corona in the open geometry of a thin accretion disk. Assuming $f_c=0.5$ and $\tau=0.5$, the two-phase thin disk-corona model produces  a cutoff power-law X-ray spectrum with $\Gamma=1.9$ and $kT_e=80\,$keV. The derived rest-frame $2-10$ keV bolometric correction, $k_{2-10}\equiv L_{\rm bol}/L_{2-10}=16$ (for an assumed UV soft photon input at 100 eV), is comparable to the value observed in low-luminosity, ``normal" type 1 AGNs \citep{Duras2020}. In the case of a corona deep in the funnel of a supercritical disk with $m\gta 3$ and the same values of $f_c$ and $\tau$, we predict  instead $\Gamma\gta 2.8$, $kT_e\lta 40\,$keV, and a bolometric correction  of $k_{2-10}\gta 70$. We observe here that comparable values of $k_{2-10}=50-80$ are actually measured at low redshifts in near-Eddington NLS1 galaxies \citep{Vasudevan2007}. Note also the steep dependence of the X-ray bolometric correction with the strength of the isotropic soft photon field: for narrower funnels (higher supercritical accretion rates) with $m\gta 5$, our model yields $\Gamma\gta 4.0$, $kT_e\lta 30\,$keV, and $k_{2-10}\gta 1600$. For comparison, the X-ray stacking of JWST-selected, faint type 1 AGNs at $z\simeq 5$ gives $k_{2-10}>500$ \citep{Maiolino2024b}. 

\section{Discussion}

The vast majority of AGNs identified by JWST, both type 1 and type 2, are not detected even with the deepest {\it Chandra} observations, not even in stacking \citep[e.g.,][]{Maiolino2024b,Yue2024}. We have discussed the possibility that these AGNs may have intrinsically faint X-ray emission because they are fed at super-Eddington rates. Specifically, we have shown that in this case their inner hot coronas will be embedded in a funnel-like reflection geometry where most of the luminosity is generated. The nearly isotropic soft photon field will Compton cool the coronal plasma  to much lower temperatures than in the standard open geometry of a thin accretion disk, and the emitted X-ray spectrum will be extremely soft, with an X-ray bolometric correction in the rest-frame 2--10 keV luminosity that can exceed those measured in standard AGNs by one or even two orders of magnitude.

Several approximations have been adopted in our calculations. Radiation-pressure driven winds are expected to naturally arise from the innermost radii of supercritical disks, carrying away mechanical energy and modifying the emission properties of the funnel regions. The accretion flows discussed here, however, are only mildly super-Eddington, with $\dot M/\dot M_{\rm Edd}$ in the range 7--12.5. In this regime, numerical simulations of  accretion flows around non-spinning MBHs show that inner hot coronae form because of the dissipation of buoyantly rising magnetic fields above the photospheres of thick disks, funnels remain optically thin, and the kinetic luminosity of the outflows is only $\sim 10\%$ of the radiation luminosity \citep{Jiang2014,Jiang2019}. It is these and similar simulation results that have inspired the analytical study presented here. Nevertheless, while thick disk models are constructed to have normal forces at the surface in balance (see Equation \ref{eq:Ffun}), the tangential components of the absorbed radiation flux are left unbalanced, and one should account for the fact that the funnel surface layers and the corona above super-Eddington flows may not be static but  moving upward \citep{Sikora_W1981,Narayan1983}. Bulk Compton scattering could then produce an additional high-energy radiation term. Eventually, more physical Comptonization models that include an X-ray reflection component from disk material should be used for detailed comparison with the observations. Numerical simulations have also shown that, while optically thin coronal  plasma with gas temperatures $\gta 10^8\,$K is generated in the inner regions of sub-Eddington accretion disks, the fraction of dissipation in this hot component (represented by our parameter $f_c$) decreases as the mass accretion rate increases \citep{Jiang2019sE}, an effect not accounted for in our calculations.

\citet{Pacucci2024} have pointed out that super-Eddington accretion would also hide coronal X-ray emission from view at high inclination angles from the poles, and this effect may offer a complementary explanation for the X-ray weakness of many JWST-selected AGNs. 
A recent study by \citet{MaiolinoAGN} examined the X-ray properties of a large sample of 71 broad-line and narrow-line AGNs identified by JWST in the GOODS fields, revealing only four X-ray detections. Notably, two of these AGNs were observed spectroscopically due to prior X-ray detection. The low fraction of X-ray bright AGNs appears unlikely to be solely a result of geometric effects, suggesting instead an inherent X-ray weakness or a combination of contributing factors. Indeed, our model of intrinsically faint X-ray emission in  supercritical flows may not be able to explain objects like the ``dormant" MBH GN-1146115 at $z=6.67$ \citep{Juod2024}, which seems to be accreting well below the Eddington limit and is undetected in X-rays with $k_{2-10}>330$ \citep{Maiolino2024b}. Yet, the properties of this faint AGN with a high black hole-to-stellar mass ratio  may support a scenario in which short bursts of super-Eddington accretion \citep{Madau2014} have resulted in black hole overgrowth. During one of these bursts, the MBH in GN-1146115 was likely fed by a thick, radiation-supported supercritical torus like those discussed in this paper.  

We note, finally, that a scaled-down version of our scenario may also apply to black hole X-ray binaries, where sources undergoing super-Eddington outbursts have been seen to transition into a hypersoft X-ray state with photon indeces $\Gamma\gta 4.5$ \citep[e.g.,][]{Uttley2015,Jin2024}. Interestingly, recent X-ray polarization observations of the supercritical source Cygnus X-3 by IXPE suggest the presence of a narrow funnel that collimates radiation emitted by the accretion disk and obscures the primary source from view \citep{Veledina2024}.

\section*{Acknowledgements}

Support for this work was provided by NASA through grant TCAN 80NSSC21K027, by grant NSF PHY-2309135 to the Kavli Institute for Theoretical Physics (KITP), and by the Italian Ministry for Research and University (MUR) under Grant `Progetto Dipartimenti di Eccellenza 2023-2027' (BiCoQ). One of us (PM) acknowledges useful discussions and inputs on this project with O. Blaes, A. Ferrara, R. Maiolino, F. Pacucci, and J. Silk. He is also grateful to V. Bromm, B. Robertson, R. Schneider, and R. Somerville for organizing the KITP workshop ``Cosmic Origins: The First Billion Years”, which provided the initial motivation for this work.

\bibliographystyle{apj}
\bibliography{paper}

\label{lastpage}

\end{document}